\documentclass[usenatbib,letterpaper]{mn2e}
\usepackage[totalwidth=480pt,totalheight=680pt,twoside=false,voffset=.5cm]{geometry}

\usepackage{graphics,graphicx,epsfig,ulem,multirow,appendix,lscape,amsmath,amssymb,color,url,float}

\begin{document}

\title[Orbital constraints on imaged companions]
  {Constraining the orbits of sub-stellar companions imaged over short orbital arcs}
\author[T. D. Pearce, M. C. Wyatt \& G. M. Kennedy]
  {Tim D. Pearce\thanks{tdpearce@ast.cam.ac.uk}, Mark C. Wyatt and Grant M. Kennedy\\
  Institute of Astronomy, University of Cambridge, Madingley Road, Cambridge, CB3 0HA, UK}
\date{Released 2002 Xxxxx XX}

\pagerange{\pageref{firstpage}--\pageref{lastpage}} \pubyear{2002}

\def\LaTeX{L\kern-.36em\raise.3ex\hbox{a}\kern-.15em
    T\kern-.1667em\lower.7ex\hbox{E}\kern-.125emX}

\newtheorem{theorem}{Theorem}[section]

\label{firstpage}

\maketitle


\begin{abstract}              
\noindent Imaging a star's companion at multiple epochs over a short orbital arc provides only four of the six coordinates required for a unique orbital solution. Probability distributions of possible solutions are commonly generated by Monte Carlo (MCMC) analysis, but these are biased by priors and may not probe the full parameter space. We suggest alternative methods to characterise possible orbits, which compliment the MCMC technique. Firstly the allowed ranges of orbital elements are prior-independent, and we provide means to calculate these ranges without numerical analyses. Hence several interesting constraints (including whether a companion even can be bound, its minimum possible semi-major axis and its minimum eccentricity) may be quickly computed using our relations as soon as orbital motion is detected. We also suggest an alternative to posterior probability distributions as a means to present possible orbital elements, namely contour plots of elements as functions of line of sight coordinates. These plots are prior-independent, readily show degeneracies between elements and allow readers to extract orbital solutions themselves. This approach is particularly useful when there are other constraints on the geometry, for example if a companion's orbit is assumed to be aligned with a disc. As examples we apply our methods to several imaged sub-stellar companions including Fomalhaut b, and for the latter object we show how different origin hypotheses affect its possible orbital solutions. We also examine visual companions of A- and G-type main sequence stars in the Washington Double Star Catalogue, and show that $\gtrsim 50$ per cent must be unbound.
\end{abstract}



\section{Introduction}
\label{sec: Introduction}

\noindent Direct imaging is now a well established technique for detecting sub-stellar companions, responsible for the discovery of at least fifty such such objects (exoplanet.eu; \citealt{Schneider11}). The required contrast sensitivity favours detection of companions far from their host stars, thus complementing the radial velocity and transit methods that are more sensitive to smaller orbital separations. However an advantage of the latter techniques is their ability to measure some orbital elements uniquely. Very often such constraints are unavailable for imaged companions; their long periods mean observations typically cover a small fraction of their orbit, so often the only known kinematic quantities are their instantaneous sky plane position and velocity (e.g. \citealt{Biller10, Neuhauser11}). The line of sight coordinates are unknown, so it is impossible to state with certainty that a companion is even bound, let alone find a unique orbital solution. Instead samples of possible orbital elements are generated, typically using a Markov chain Monte Carlo (MCMC) analysis to find possible orbital fits to the observations (e.g. \citealt{Ford06, Chauvin12, Kalas13, Beust14, Pueyo14}). This method has many advantages; it easily incorporates observational uncertainties, and can fit orbits using a wide range of observational constraints (such as additional radial velocity measurements, e.g. \citealt{Crepp12}). However an issue arises when considering the resulting samples of possible orbits, as it is assumed that these samples represent the probability distributions associated with the orbital elements. Whether they actually do is unclear, as the arbitrary choice of prior probability distributions (hereafter ``priors'') associated with the unknown quantities influences the outcomes of the MCMC; given the same observables, different priors will produce different results.

In this paper we argue that, for companions imaged over short orbital arcs, a good alternative to MCMC is to show how orbital elements vary with the assumed line of sight coordinates. This method has previously been used by \citet{Golimowski98}, \citet{Golimowski00} and \citet{Hinkley10} to characterise the semi-major axis and eccentricity of a companion; extending it to all orbital elements would allow a reader to extract full orbital solutions themselves. It is also prior-independent. We also provide general, prior-independent bounds on the orbital elements of an imaged companion. Both methods are simple to use and may be applied as soon as orbital motion is detected, and provide complimentary information to the MCMC technique.

The equations and techniques presented in this paper are general, and throughout the paper we apply the methods to Fomalhaut b as an example (a companion which clearly shows linear motion relative to its host, \citealt{Kalas13}). The layout of this paper is as follows. In Section \ref{sec: parameters} we outline the problem, and in Section \ref{sec: bound_circ_criteria} we provide a simple method to identify whether a general imaged companion can possibly be bound. We discuss difficulties in interpreting generated distributions of possible orbits in Section \ref{sec: element_distributions}, and how these may be overcome. In Section \ref{sec: distribution_bounds} we provide general, prior-independent limits on orbital elements. We provide some example applications of our methods in Section \ref{sec: application}, and in Section \ref{sec: physically_motivated_priors} we discuss how the bounds on Fomalhaut b's orbit change when various physically motivated priors are considered. In Section \ref{sec: limitations} we discuss the limitations of our method.


\section{Known and unknown parameters}
\label{sec: parameters}

\noindent Consider a binary system, comprising a primary and a companion. We assume the total binary mass $M$ is known, and we define $\mu \equiv G M$ where $G$ is the gravitational constant. We assume that multi-epoch imaging has been performed over a short orbital arc, yielding three more parameters: the binary's projected angular separation $R_{\rm ang}$, the angular sky plane velocity $V_{\rm ang}$ of one component relative to the other, and the angle $\varphi$ between these projected separation and velocity vectors. Assuming the Earth-binary distance $d$ is known, we can convert $R_{\rm ang}$ and $V_{\rm ang}$ into projected distance $R$ and velocity $V$. Finally we assume no orbital curvature is observed. See Appendix \ref{app: RV_from_obs} for the derivation of $R$, $V$ and $\varphi$ from observational data.

\begin{figure}
  \centering
      \includegraphics[width=8cm]{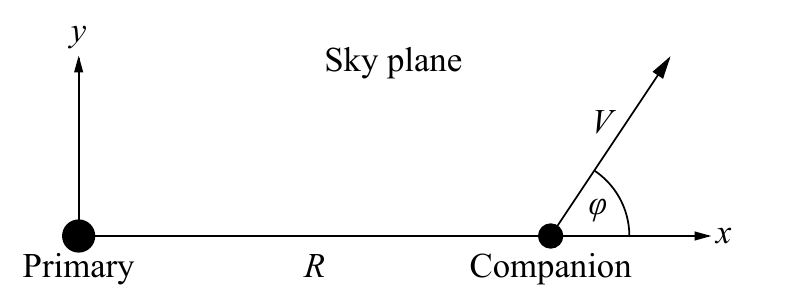}
  \caption{The coordinate system used in this paper, where the ($x$, $y$) plane is the sky plane. $R$ and $V$ are the projected sky plane position and speed of the companion relative to the primary, and $\varphi$ the angle between these. The $z$ direction is chosen such that $0 \leq \varphi \leq 180^\circ$, so here points out of the page (towards the reader).}
  \label{fig: coordinate_system}
\end{figure}

We define a coordinate system centred on the primary with the $x$, $y$ plane defining the sky plane and the projected separation vector lying along the $x$ axis (Figure \ref{fig: coordinate_system}). For simplicity we define $\varphi$ to lie in the range $0 \leq \varphi \leq 180^\circ$, thus setting whether the observer lies at positive or negative $z$ (this will affect angular orbital elements, as described in Section \ref{sec: distribution_bounds}). Thus we have two unknowns, the companion's position $z$ and velocity $\dot{z}$ perpendicular to the sky plane, for which we have no priors and may only limit by assuming that the binary is bound.

We define the positive dimensionless parameter $B$, which is a combination of observable (and known) quantities

\begin{equation}
B \equiv \frac{V^2 R}{2\mu}. 
\label{eq: B}
\end{equation}

\noindent Equivalently

\begin{equation}
B \equiv \frac{1}{8\pi^2}\left(\frac{d}{{\rm pc}}\right)^3\left(\frac{V_{\rm ang}}{{\rm arcsec / yr}}\right)^2\left(\frac{R_{\rm ang}}{{\rm arcsec}}\right)\left(\frac{M}{{\rm M}_\odot}\right)^{-1}.
\label{eq: B observables}
\end{equation}

\noindent This parameter will prove useful throughout the paper.


\section{Criterion for a bound orbit}
\label{sec: bound_circ_criteria}

\noindent Consider a binary with one component at instantaneous position $\mathbf{r} = (R,0,z)$ and velocity $\mathbf{v} = (V\cos\varphi,V\sin\varphi,\dot{z})$ relative to the other. The two objects may only be bound if $v$ is below the escape speed, i.e. $v^2 r < 2\mu$ (where $r \equiv |\mathbf{r}|$ and $v \equiv |\mathbf{v}|$). The minimum possible values of $v$ and $r$ occur if $z = \dot{z} = 0$, hence a system may only be bound if its sky plane coordinates satisfy

\begin{equation}
 B < 1.
\label{eq: bound}
\end{equation}

\noindent Hence Equations \ref{eq: B} and \ref{eq: bound} may be used to vet common proper motion companions for possible physical association (see Section \ref{sec: application}). It follows that for bound orbits

\begin{equation}
\left|\frac{z}{R}\right| < \sqrt{B^{-2} -1} {\text {\;\;\;\; and \;\;\;\;}} \left|\frac{\dot{z}}{V}\right| < \sqrt{B^{-1} -1}.
\label{eq: z and vz max}
\end{equation}

\noindent Note that unbound solutions are always allowed, as there are no practical upper limits on possible values of $|z|$ and $|\dot{z}|$.


\section{Distributions of possible orbital elements}
\label{sec: element_distributions}

\noindent We now consider the companion's orbital elements. A Keplerian orbit is completely described by five elements: semi-major axis $a$, eccentricity $e$, inclination to a reference plane $i$, longitude of ascending node $\Omega$ (from a reference direction) and argument of pericentre $\omega$. Additionally the true anomaly $f$ determines the body's location along its orbit. We define the $x$, $y$ (sky) plane as the reference plane and $x$ (the primary-companion separation vector) as the reference direction. Appendix \ref{app: element_equations} shows how these orbital elements are derived from a companion's instantaneous position and velocity. Importantly, all six elements require the companion's six-dimensional coordinates to be known for unique determination. Hence given a companion's sky plane coordinates, we must assume values of $z$ and $\dot{z}$ to generate orbital elements. Varying these assumed values generates various orbital solutions, given the observed parameters.

\begin{figure*}
  \centering
      \includegraphics[width=16cm]{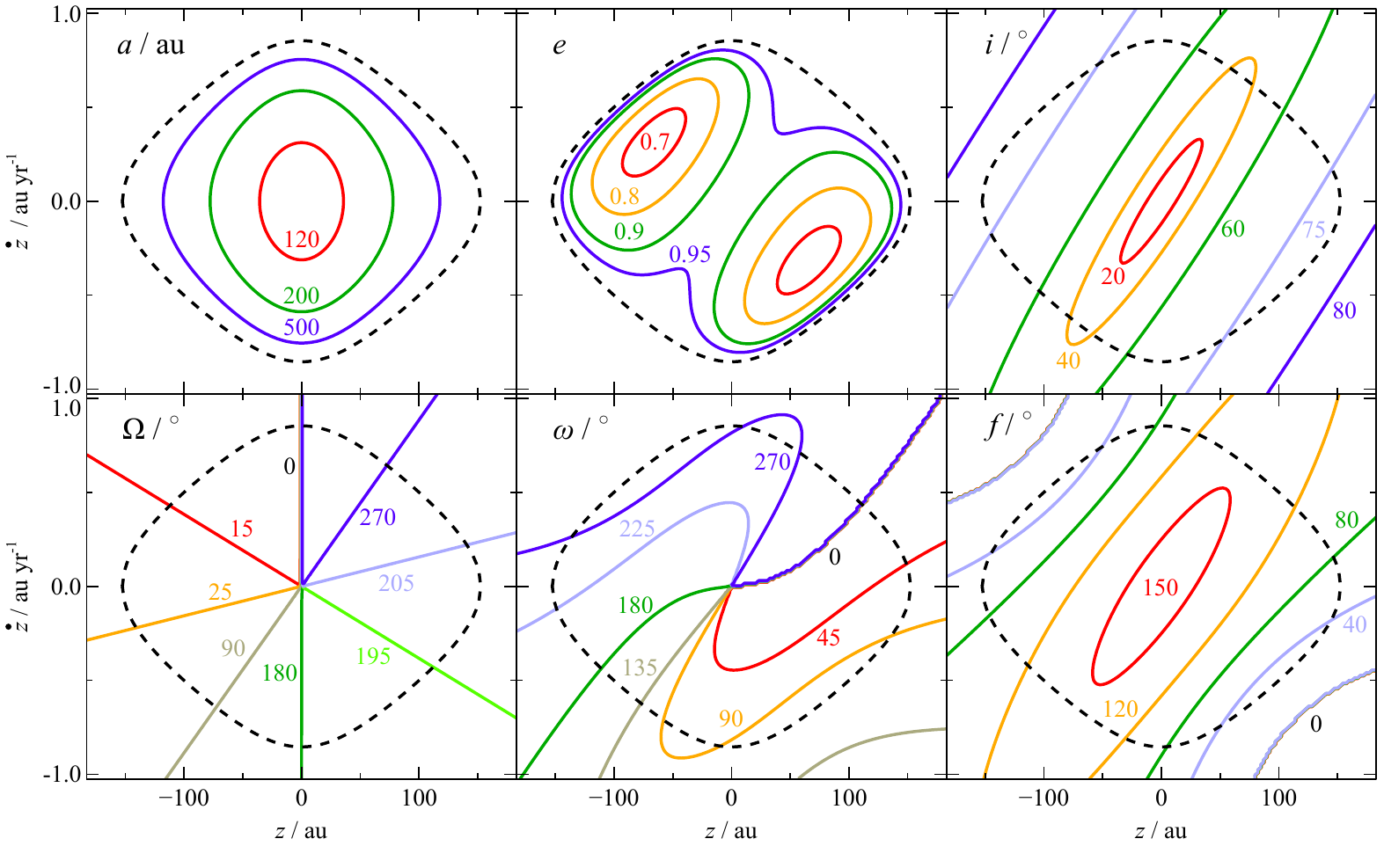}
  \caption{The dependence of an imaged companion's orbital elements on its unknown line of sight position $z$ and velocity $\dot{z}$. This example is Fomalhaut b, which has $R_{\rm ang} = 12.54$ arcsec, $V_{\rm ang} = 0.119$ arcsec/yr, $\varphi = 21^\circ$ and $M_* = 1.92 M_\odot$. Solutions outside the dashed lines are unbound; note that $a \rightarrow \infty$ and $e \rightarrow 1$ as the orbit approaches the unbound limit. The plot does not include uncertainties on $R$, $V$, $\varphi$ and $M_*$, which would broaden the contour lines. For Fomalhaut's observational errors, the $1\sigma$ uncertainty on the $z$ value corresponding to each contour is of order $\pm 10$ au, and that on $\dot{z}$ is of order $\pm 10^{-2}$ au yr$^{-1}$. Plots like this allow the reader to extract possible orbits themselves, by choosing a combination of $z$ and $\dot{z}$ and reading off the corresponding orbital elements from each panel.}
  \label{fig: element_contours}
\end{figure*}

To demonstrate this we require an example companion which shows linear motion over a short orbital arc, and we will use Fomalhaut b. We also use this object as our primary example for the remainder of the paper. Fomalhaut b is a $\lesssim$ Jupiter mass object orbiting a $1.92 M_\odot$ star $7.7$ pc from Earth \citep{Kalas08, Janson12, Mamajek12}. \cite{Kalas13} measured its position at four epochs between 2004 and 2012, detecting orbital motion but not acceleration. We fit the observed positions with a linear trend to derive the sky plane velocity. At the first epoch $R_{\rm ang} = 12.54 \pm 0.02$ arcsec, and we fit $V_{\rm ang} = 0.119 \pm 0.006$ arcsec/yr and $\varphi = 21 \pm 2^\circ$. Neglecting the observational uncertainties, the resulting orbital elements as functions of $z$ and $\dot{z}$ are shown on Figure \ref{fig: element_contours}. The plot shows that very different orbits are possible depending on the unknown line of sight coordinates, and also shows the relationships between orbital elements as $z$ and $\dot{z}$ are varied. We will refer back to this plot throughout the paper. Note that whilst observational uncertainties were omitted for simplicity, they need not have been; we could have drawn many combinations of the observables using their measured values and uncertainties, and produced a contour plot for each combination. Superimposing these plots would accommodate the observational uncertainties by broadening the contour lines, but we choose not to do this in our example for clarity.

\begin{figure*}
  \centering
      \includegraphics[width=18cm]{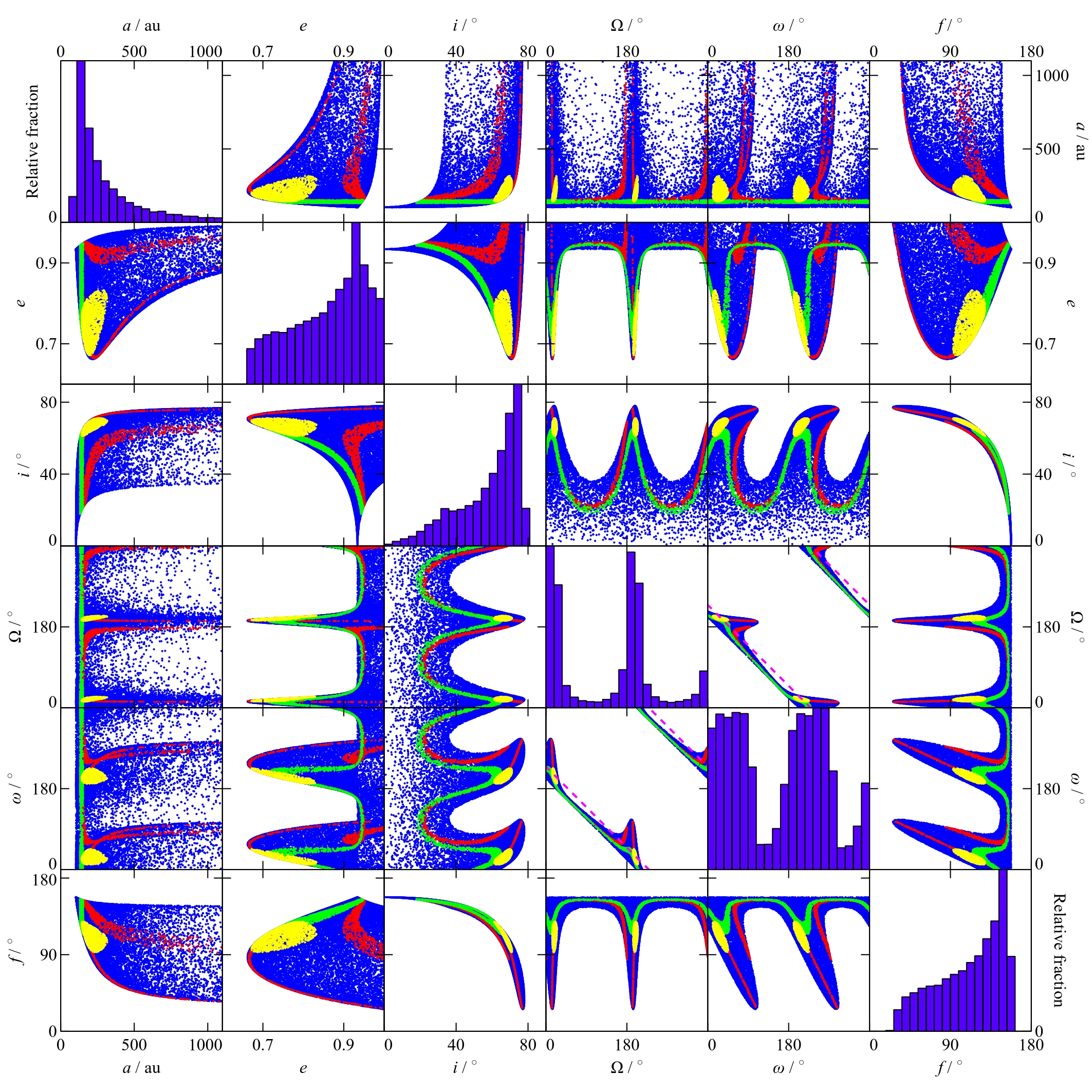}
  \caption{Possible orbital solutions for an imaged companion, showing the degeneracies between different elements. Again, Fomalhaut b is used as an example. Each point on the off-diagonal plots is a possible orbital solution. Blue solutions were generated by drawing $z$ and $\dot{z}$ uniformly, and the histograms show the distributions of possible orbital elements resulting from this prior. For any companion, different priors change the density of solutions within the bounds but \textit{not} the bounds themselves. Here observational uncertainties are omitted; these would blur the plots, but by a small amount in this case. We also show solutions for some physically motivated priors specific to Fomalhaut, as detailed in Section \ref{sec: physically_motivated_priors}; yellow points are orbits within $5^\circ$ of the system's debris disc plane, green orbits have semi-major axes within 10 au of that of the disc, and red orbits pass within 5 au of the disc when Fomalhaut b crosses it in projection in $\sim 50$ years' time. Note $a \rightarrow \infty$ as $e \rightarrow 1$, however we only show $a \leq 1100$ au for clarity. A degeneracy exists between $\Omega_{\rm disc}$, $\omega_{\rm disc}$ and $\Omega_{\rm disc} + 180^\circ$, $\omega_{\rm disc} + 180^\circ$ \citep{Beust14}, and we plot both solutions here. The dashed magenta lines on the $\Omega$ versus $\omega$ panels show $\varpi = \varpi_{\rm disc}$, where $\varpi \equiv \Omega + \omega$.}
  \label{fig: element_distributions}
\end{figure*}

Drawing many values of $z$ and $\dot{z}$, with uniform priors on both variables, results in the orbits shown as the blue points on Figure \ref{fig: element_distributions} (note only bound orbits are shown). Evidently there are many degeneracies between different elements that are apparent on these plots. The histograms show the corresponding distributions of possible orbital element values, for the assumption of uniform priors. Again, we have neglected observational uncertainties when generating this plot; their inclusion would broaden the distributions.

\begin{figure}
  \centering
      \includegraphics[width=8cm]{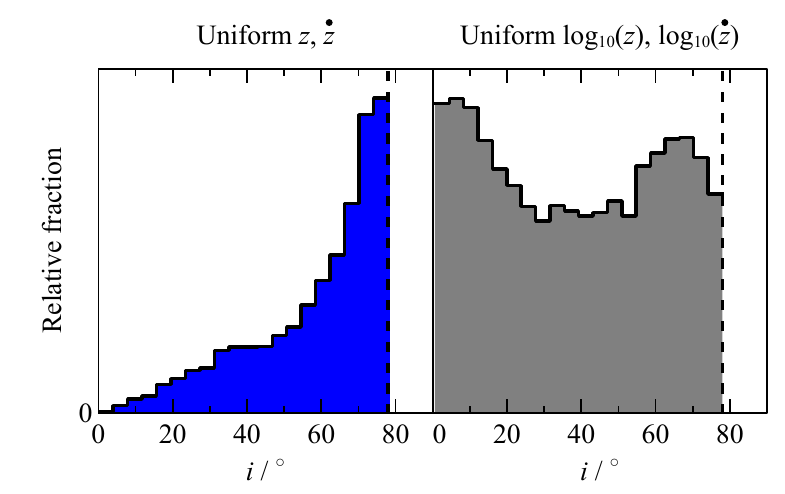}
  \caption{The effect of different priors on the distribution of orbital inclination solutions for Fomalhaut b (if bound). The left plot shows the distribution found using a uniform prior on $z$ and $\dot{z}$. The right plot shows uniform logarithmic priors, where $\log_{10}(z/R)$ and $\log_{10}(\dot{z}/V)$ are drawn between 0.01 and the maximum values from Equation \ref{eq: z and vz max}. The dashed lines show the maximum allowed inclination, using the method in Section \ref{sec: distribution_bounds}.}
  \label{fig: i_diff_priors}
\end{figure}

The histograms on Figure \ref{fig: element_distributions} agree with distributions of possible orbital elements obtained using MCMC \citep{Kalas13,Beust14}. However consideration of Figure \ref{fig: element_contours} shows that care must be taken when interpreting such histograms. Firstly, if these distributions are interpreted as probabilities then it is important to remember that these depend on the assumed priors. For example, had we chosen a prior which biased towards low $|z|$ and $|\dot{z}|$ (such as uniform $\log |z|$ and $\log |\dot{z}|$) then a low inclination would be favoured, the opposite of the high inclination inferred from uniform $z$, $\dot{z}$ priors. We show this on Figure \ref{fig: i_diff_priors}. While uniform $z$ and $\dot{z}$ priors may seem the most unbiased priors that can be assumed, since it implies no knowledge of these variables, this ignorance could be applied in other ways. For example assuming ignorance of the orbital properties, say through a uniform prior on eccentricity, will result in a different set of solutions to that calculated assuming uniform $z$ and $\dot{z}$. Hence the ``most likely'' orbital elements clearly depend on the assumed priors. This is further explored in Section \ref{sec: physically_motivated_priors}, where we discuss physically motivated priors that can be applied to Fomalhaut b.

Secondly, distributions of possible orbital elements alone say nothing about element degeneracies. For example, the semi-major axis and inclination histograms for uniform $z$ and $\dot{z}$ priors on Figure \ref{fig: element_distributions} imply that the most likely orbit has $a \sim 150$ au and $i \sim 70^\circ$. However the $a$ versus $i$ plot on Figure \ref{fig: element_distributions} shows that no solutions exist with \textit{both} $a = 150$ au and $i = 70^\circ$. This is not a problem if one has access to the list of orbital solutions used to create the histograms, but great care must be taken if attempting to interpret possible orbits using histograms alone.

We suggest two methods to characterise possible orbits as alternatives to MCMC, which unlike the latter technique are both prior-independent and probe the full region of allowed parameter space. Firstly, whilst histograms of possible elements are prior-dependent regardless of the method used to generate them, the allowed ranges of orbital elements are set by the known parameters $R$, $V$, $\varphi$ and $\mu$. For example, re-generating Figure \ref{fig: element_distributions} using different $z$ and $\dot{z}$ priors produces the same bounds, but the density of solutions changes (see Figure \ref{fig: i_diff_priors}). These bounds can be calculated (see Section \ref{sec: distribution_bounds}), so we suggest that such bounds are quoted as a prior-independent alternative to histograms of possible solutions.

Secondly, the degeneracies between elements are also prior-independent. Figure \ref{fig: element_contours} fully describes these degeneracies, as each combination of $z$ and $\dot{z}$ corresponds to a single set of orbital solutions. This figure also allows possible orbits to be extracted by the reader, without having to publish lists of solutions (as would be required with the MCMC method). Hence a poorly sampled orbit may be defined by contour plots of the orbital elements corresponding to different $z$ and $\dot{z}$ combinations (as on Figure \ref{fig: element_contours}), as well as quoting the aforementioned bounds. These two methods would compliment the MCMC method, and may even be more informative than the latter (at least when motion is only detected over a short portion of the orbit, see Section \ref{sec: limitations}).


\section{Orbital element bounds}
\label{sec: distribution_bounds}

\noindent We now provide general bounds for several orbital elements, which may be quickly calculated as soon as orbital motion is observed without the need for numerical analyses. Figures \ref{fig: element_contours} and \ref{fig: element_distributions} show that $\Omega$ and $\omega$ may assume any value between 0 and $360^\circ$, and this holds for any combination of sky plane coordinates. However $e$ has a clear minimum, and assuming the companion is bound also limits the values of $a$, $i$ and sometimes $f$. We provide the extreme values of these elements for any bound companion imaged over a short orbital arc (note that defining $0 \leq \varphi \leq 180^\circ$ also sets $0 \leq i \leq 90^\circ$ and affects $\Omega$ and $\omega$, but makes no difference to $a$, $e$ and $f$).

Firstly examination of Equation \ref{eq: a} shows that the minimum semi-major axis for a bound object will always occur if $z = \dot{z} = 0$, i.e.

\begin{equation}
\frac{a_{\rm min}}{R} = \frac{1}{2}(1 - B)^{-1}.
\label{eq: min a}
\end{equation}

\noindent No upper limit may be placed on $a$, as it tends to infinity as the orbit approaches the unbound limit.

\begin{figure}
  \centering
      \includegraphics[width=8cm]{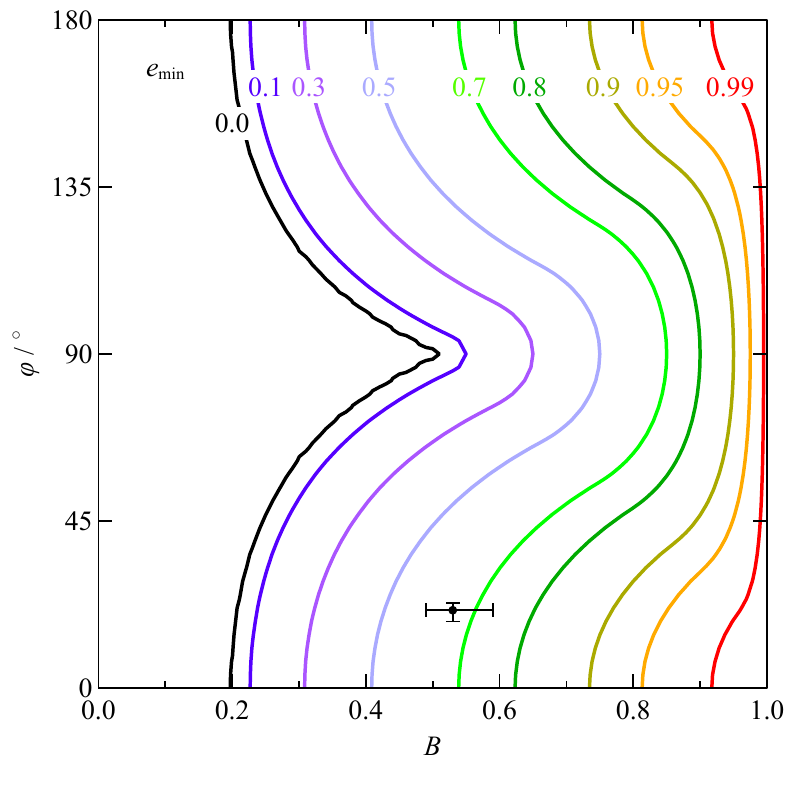}
  \caption{The minimum possible eccentricity of a companion as a function of its sky plane coordinates. The contours were found numerically by varying the line of sight position and velocity at each point. This plot is completely general, and the point shows $B$ and $\varphi$ for Fomalhaut b as an example.}
  \label{fig: emins}
\end{figure}

No analytic solution for the minimum eccentricity in terms of $z$ and $\dot{z}$ can be found, so we calculate this numerically instead. Firstly we reduce the number of parameters by making Equations \ref{eq: a} to \ref{eq: rdot} dimensionless (i.e. dividing all lengths and velocities by $R$ and $V$ respectively). This yields a natural coupling between the variables $R$, $V$ and $\mu$ in terms of $B$; thus we reduce the number of parameters from six ($R$, $V$, $\varphi$, $\mu$, $z$ and $\dot{z}$) to four ($B$, $\varphi$, $z/R$ and $\dot{z}/V$). In Appendix \ref{app: element_equations} we recast the eccentricity equation (Equation \ref{eq: e}) in terms of these dimensionless variables (Equation \ref{eq: e full}). The minimum eccentricity will occur at a certain combination of $z/R$ and $\dot{z}/V$, hence its value is only a function of $B$ and $\varphi$. On Figure \ref{fig: emins} we plot the minimum eccentricity contours for each combination of $B$ and $\varphi$, found by numerically varying $z/R$ and $\dot{z}/V$. This plot is completely general for any system imaged over a short orbital arc.

A numerical fit to the $e_{\rm min} = 0$ contour on Figure \ref{fig: emins} shows that a companion cannot have a circular orbit unless

\begin{equation}
B \lesssim 0.2 + 0.3\exp(-0.04|\varphi - 90^\circ|),
\label{eq: circ_criterion}
\end{equation}

\noindent where $\varphi$ is in degrees. In particular note that the eccentricity cannot be zero if $B > 0.5$, as for a circular orbit $v^2 r / \mu = 1$ (recalling that $r$ and $v$ are the relative 3-dimensional position and velocity respectively). There is no upper limit on eccentricity.

\begin{figure}
  \centering
      \includegraphics[width=8cm]{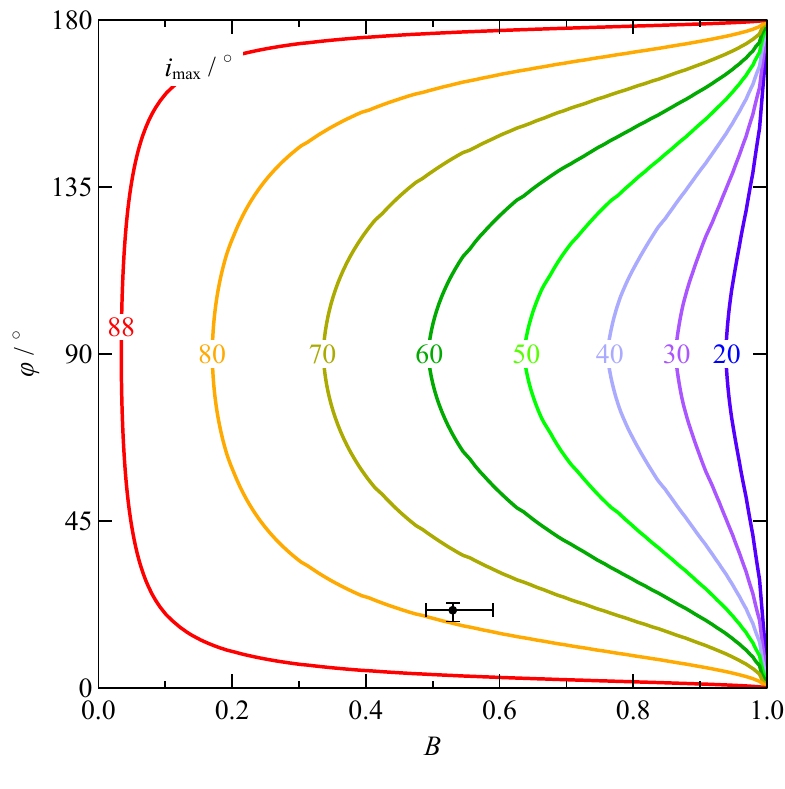}
  \caption{A bound companion's maximum possible inclination to the sky plane, as a function of its sky coordinates. This plot is completely general. Since we define $0^\circ \leq \varphi \leq 180^\circ$, $i$ is confined to the range $0^\circ \leq i \leq 90^\circ$. Note that if Equation \ref{eq: bound} is satisfied, $i = 0^\circ$ is always a possible bound solution if $\varphi \neq 0^\circ$ or $180^\circ$. The point shows $B$ and $\varphi$ for Fomalhaut b.}
  \label{fig: imaxs}
\end{figure}

We now bound the companion's inclination. By definition this equals $0^\circ$ if $z=\dot{z}=0$, and this solution is always allowed (unless $\varphi = 0^\circ$ or $180^\circ$, in which case the inclination must be $90^\circ$ if the object is bound). The inclination of bound orbits also has a maximum, again set by $B$ and $\varphi$, and we plot these contours on Figure \ref{fig: imaxs}. This maximum occurs if the companion is just bound, i.e. at $e \approx 1$. This is because the magnitude of $\cos i$ is set by $h \equiv |\mathbf{r} \times \mathbf{v}|$ (Equation \ref{eq: i} - noting that $\mathbf{h}.\mathbf{\hat{z}}$ is an observable and thus independent of $z$ and $\dot{z}$), which is minimal at $z=\dot{z}=0$ and forever increases as $|z|$ and $|\dot{z}|$ increase.

\begin{figure}
  \centering
      \includegraphics[width=8cm]{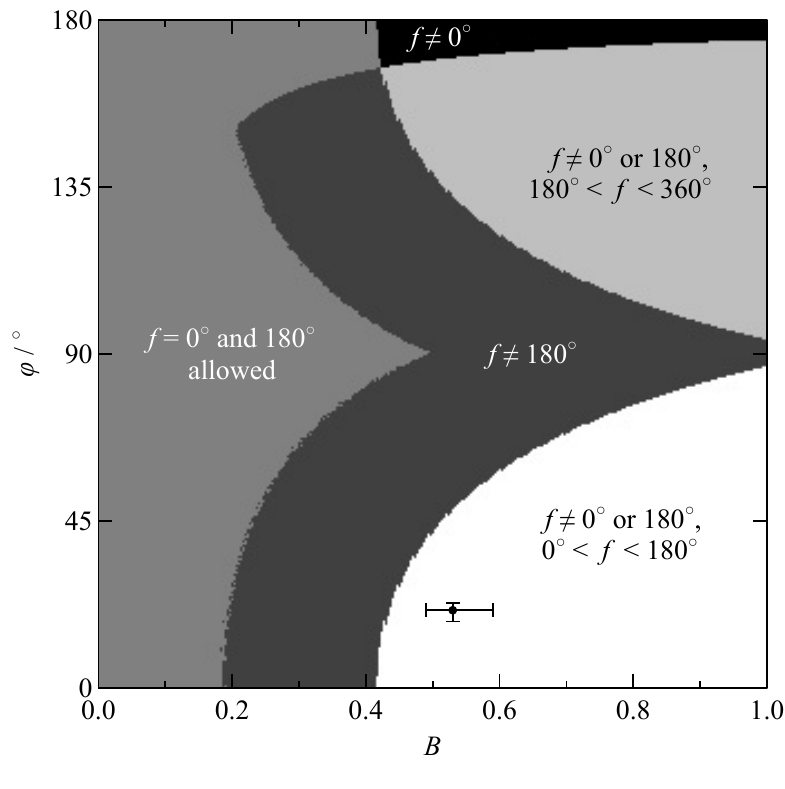}
  \caption{General constraints on a bound companion's true anomaly $f$. A bound companion in the $f \neq 180^\circ$ region cannot be at apocentre but may be at pericentre, and vice-versa for the $f \neq 0^\circ$ region. If a companion cannot be at apocentre or pericentre, its true anomaly is further constrained inside $0^\circ < f < 180^\circ$ or $180^\circ < f < 360^\circ$ (note that the actual range of allowed $f$ values will lie within these bounds; for example the point shows Fomalhaut b, which actually has $25^\circ \leq f \leq 133^\circ$).}
  \label{fig: apo_peri_allowed}
\end{figure}

The true anomaly is sometimes constrained for a bound orbit (as it is for Fomalhaut b). Of particular interest is whether an imaged companion could be near pericentre or apocentre. We show this on Figure \ref{fig: apo_peri_allowed}, as a general function of $B$ and $\varphi$. If $B \lesssim 0.2$ then a bound companion may have any true anomaly; this is because $e_{\rm min} \approx 0$ (Figure \ref{fig: emins}), so all pericentre orientations (and hence true anomalies) are energetically allowed. For this reason, the region of $B$ and $\varphi$ space where all $f$ values are allowed is very similar to the $e_{\rm min} < 0.1$ region on Figure \ref{fig: emins}. If $B \gtrsim 0.2$ then, depending on $\varphi$, it may not be possible for a bound companion have certain values of $f$. We highlight regions of $B$, $\varphi$ space for which a bound companion cannot be at pericentre, apocentre or either; in the latter case, it is also known whether the companion is moving towards pericentre or apocentre. This is because if the companion is moving away from the star in projection ($0^\circ \leq \varphi < 90^\circ$), it would be moving from pericentre to apocentre if its orbit lay in the sky plane. If it is actually moving from apocentre to pericentre, its line of sight position / velocity must be reasonably high. However these required line of sight coordinates may be too large for the orbit to be bound, particularly if $B$ is large, in which case it is not possible for the companion to be moving towards pericentre and be bound. The opposite argument is true if $90^\circ < \varphi \leq 180^\circ$.

\begin{figure}
  \centering
      \includegraphics[width=8cm]{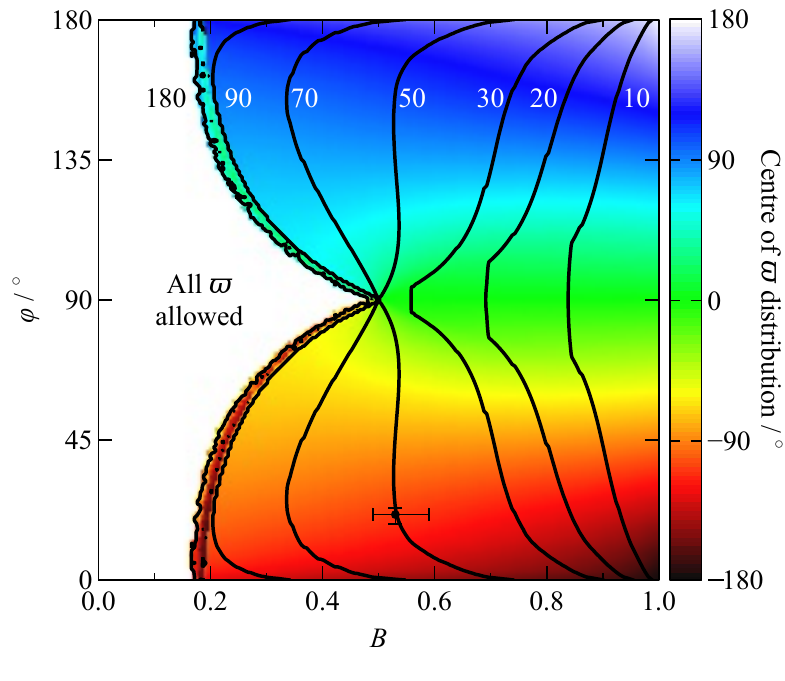}
  \caption{Allowed values of a general companion's longitude of pericentre $\varpi \equiv \Omega + \omega$. For each combination of $B$ and $\varphi$ there are a range of possible values of $\varpi$. Colours show the value of $\varpi$ in the middle of the allowed range, i.e. $[\max(\varpi) + \min(\varpi)]/2$. Contours show the half width of the range, i.e. $[\max(\varpi) - \min(\varpi)]/2$. Much of the region between the $90^\circ$ and $180^\circ$ contours is dominated by numerical noise and should be ignored. The point shows Fomalhaut b, which has $203^\circ \leq \varpi \leq 304^\circ$ (i.e. $-157^\circ \leq \varpi \leq -124^\circ$ on the plot).}
  \label{fig: cpi_range}
\end{figure}

Finally although $\Omega$ and $\omega$ are always unconstrained, for many combinations of $B$ and $\varphi$ the companion's longitude of pericenter $\varpi \equiv \Omega + \omega$ is bounded reasonably well. For many imaged companions it should thus be possible to identify the orientation of pericentre on the sky, which may be interesting if other bodies have been detected in the system. On Figure \ref{fig: cpi_range} we show the value of $\varpi$ at the centre of its allowed range, and also the half width of the range. The range of allowed $\varpi$ values decreases as $B$ approaches 1; this is because the ranges of $z$ and $\dot{z}$ resulting in a bound orbit also decreases as $B \rightarrow 1$. For an companion with $B \approx 1$ to be bound, $z$ and $\dot{z}$ must be approximately zero; in this case it can be shown that $\varpi \rightarrow 2\varphi - 180^\circ$. For example it is clear from Figure \ref{fig: cpi_range} that the projected location of Fomalhaut b's pericentre (if bound) must lie on the opposite side of the star to its current projected location, and this will prove important in Section \ref{sec: physically_motivated_priors}.

\section{Application to imaged companions}
\label{sec: application}

\noindent Our methods may be applied as soon as a companion's orbital motion is detected, firstly to establish whether it could be bound and then to constrain its semi-major axis, eccentricity, inclination and true anomaly. Even broad constraints are interesting; a non-zero minimum eccentricity may point towards a certain formation mechanism or past dynamical evolution (e.g. \citealt{Mayer04}, \citealt{Chatterjee08}). An unseen mass can make a companion's eccentricity appear greater than it really is \citep{Pearce14}, hence a high minimum eccentricity could justify a search for other bodies in the system. Additionally were a companion's maximum inclination incompatible with that derived for another component (e.g. a resolved debris disc or the stellar rotation axis), one could infer previous scattering or Kozai evolution even if the exact orbit were unknown (e.g. \citealt{Ford00, Greaves14}).

We found the extreme orbital elements of several imaged sub-stellar companions as examples, listed in Table \ref{tab: example_systems}. Each shows linear or near-linear motion relative to its host, which is compatible with a bound orbit ($B < 1$). For each example system we drew $10^4$ combinations of the astrometry, mass and distance measurements, assuming Gaussian uncertainties on these values. For each combination we then derived the corresponding values of $R$, $V$, $\varphi$ and $B$, and also the extreme orbital elements ($a_{\rm min}$ etc.). Table \ref{tab: example_systems} shows the median values of the resulting distributions and the $1\sigma$ uncertainties. The uncertainties on the extreme orbital elements are generally large, as the sky motions are typically poorly constrained. Nonetheless Fomalhaut b and PZ Tel B (if bound) must be highly eccentric. Furthermore a minimum eccentricity compatible with zero is still useful for modelling purposes, even if the uncertainties are large. Note that we drew $10^4$ values of the observables to demonstrate the uncertainties on the derived quantities; alternatively these errors could be roughly estimated by propagating the uncertainties on $B$ and $\varphi$ through Equation \ref{eq: min a}, and placing $B$ and $\varphi$ points with error bars on Figures \ref{fig: emins} and \ref{fig: imaxs} (as shown for Fomalhaut b on these figures). 

\renewcommand{\tabcolsep}{3pt}
\begin{center}
\begin{table*}
\begin{tabular}{c c c c c c c c c c c}

\hline																																																					
Companion	&	$d$ / pc			&	$M_*$ / $M_\odot$			&	$R$ / au					&	$V$  / au yr$^{-1}$					&	$\varphi$ / $^\circ$					&	$B$					&	$\min(a)$ / au					&	$\min(e)$					&	$\max(i)$ / $^\circ$					&	Ref.	\\
\hline																																																					
2M 0103(AB) b	& $	47	\pm	3	$ & $	0.36	\pm	0.02	$ & $^\dagger	82	^{+	5	}_{-	6	}$ & $^\dagger	0.36	^{+	0.08	}_{-	0.07	}$ & $	46	^{+	22	}_{-	7	}$ & $	0.26	^{+	0.3	}_{-	0.03	}$ & $	55	^{+	40	}_{-	4	}$ & $	0.0	^{+	0.6	}_{-	0.0	}$&$	76	^{+	2	}_{-	10	}$&	1	\\
Fomalhaut b	& $	7.70	\pm	0.04	$ & $	1.92	\pm	0.02	$ & $	96.6	^{+	0.5	}_{-	0.6	}$ & $	0.91	^{+	0.05	}_{-	0.04	}$ & $	21	^{+	2	}_{-	3	}$ & $	0.53	^{+	0.06	}_{-	0.04	}$ & $	101	^{+	18	}_{-	6	}$ & $	0.66	^{+	0.07	}_{-	0.06	}$&$	78	^{+	2	}_{-	3	}$&	2	\\
GJ 504 b	& $	17.56	\pm	0.08	$ & $	1.22	\pm	0.08	$ & $	43.6	^{+	0.2	}_{-	0.3	}$ & $	0.9	^{+	0.2	}_{-	0.3	}$ & $	85	^{+	15	}_{-	20	}$ & $	0.3	^{+	0.3	}_{-	0.1	}$ & $	29	^{+	26	}_{-	1	}$ & $	0.0	^{+	0.4	}_{-	0.0	}$&$	74	^{+	6	}_{-	19	}$&	3	\\
Gliese 229 B	& $	5.77	\pm	0.04	$ & $	0.570	\pm	0.002	$ & $	44.8	\pm	0.3			$ & $	0.5	\pm	0.1			$ & $	144	^{+	20	}_{-	8	}$ & $	0.3	^{+	0.2	}_{-	0.1	}$ & $	30	^{+	10	}_{-	2	}$ & $	0.0	^{+	0.5	}_{-	0.0	}$&$	84	^{+	3	}_{-	11	}$&	4	\\
GQ Lup B	& $	140	\pm	50	$ & $	0.7	\pm	0.1	$ & $	103	^{+	32	}_{-	37	}$ & $	0.4	\pm	0.1			$ & $	125	^{+	11	}_{-	7	}$ & $	0.0	^{+	0.5	}_{-	0.0	}$ & $	47	^{+	92	}_{-	4	}$ & $	0.0	^{+	0.6	}_{-	0.0	}$&$	88	^{+	2	}_{-	27	}$&	5	\\
HR 8799 b	& $	39	\pm	1	$ & $	1.5	\pm	0.3	$ & $	68	^{+	1	}_{-	2	}$ & $	0.92	^{+	0.02	}_{-	0.05	}$ & $	95	\pm	2			$ & $	0.44	^{+	0.2	}_{-	0.04	}$ & $	57	^{+	40	}_{-	10	}$ & $	0.0	^{+	0.3	}_{-	0.0	}$&$	64	^{+	3	}_{-	9	}$&	6	\\
HR 8799 e	& $	39	\pm	1	$ & $	1.5	\pm	0.3	$ & $	14.4	\pm	0.4			$ & $	2.0	^{+	0.2	}_{-	0.3	}$ & $	98	^{+	6	}_{-	6	}$ & $	0.40	^{+	0.2	}_{-	0.06	}$ & $	11	^{+	7	}_{-	1	}$ & $	0.0	^{+	0.4	}_{-	0.0	}$&$	67	^{+	3	}_{-	15	}$&	6	\\
PZ Tel B	& $	52	\pm	3	$ & $	1.13	\pm	0.03	$ & $	13.4	^{+	0.7	}_{-	0.6	}$ & $	1.7	\pm	0.1			$ & $	5	^{+	2	}_{-	2	}$ & $	0.38	^{+	0.08	}_{-	0.06	}$ & $	11	^{+	3	}_{-	1	}$ & $	0.5	^{+	0.1	}_{-	0.1	}$&$	88	\pm	1			$&	5	\\
TWA 5 B	& $	44	\pm	4	$ & $	0.7	\pm	0.1	$ & $	87	^{+	7	}_{-	9	}$ & $	0.52	\pm	0.07			$ & $	133	^{+	5	}_{-	7	}$ & $	0.4	^{+	0.2	}_{-	0.1	}$ & $	62	^{+	50	}_{-	7	}$ & $	0.0	^{+	0.6	}_{-	0.0	}$&$	75	^{+	4	}_{-	13	}$&	7	\\
\hline																																																					

\end{tabular}
\caption{Sky coordinates of several imaged sub-stellar companions, for which orbital motion but little to no curvature is detected. $V$ and $\varphi$ were derived from a linear fit to the sky positions, and $R$ is the fitted sky separation at the first observation epoch. Note inclination is only bounded by assuming the companion is bound. Errors are $1\sigma$. Observation references: (1) \citealt{Delorme13}, (2) \citealt{Kalas13}, (3) \citealt{Kuzuhara13}, (4) \citealt{Golimowski98}, (5) \citealt{Ginski14}, plus refs. therein, (6) refs. in \citealt{Gozdziewski14}, (7) \citealt{Neuhauser10}. $^\dagger$Coordinates relative to AB's barycentre. The large uncertainties for GQ Lup B result from the uncertainty on its distance. The small discrepancies between our PZ Tel B results and those of \citet{Ginski14} are likely due to the latter allowing for observations of orbital curvature, and also the reason given in Section \ref{sec: limitations}.}
\label{tab: example_systems}
\end{table*}
\end{center}

\begin{figure}
  \centering
      \includegraphics[width=8cm]{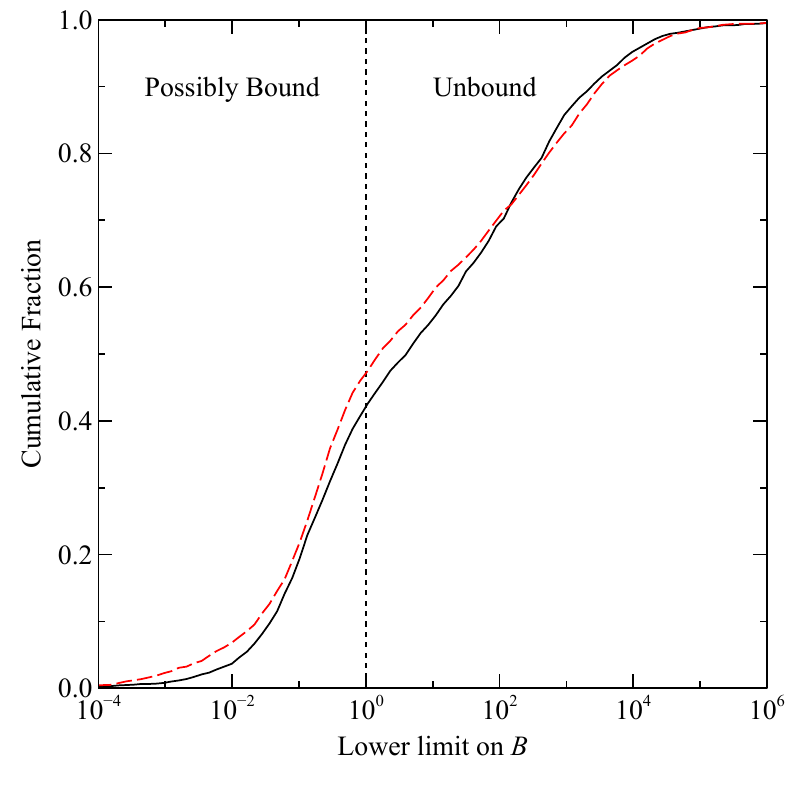}
  \caption{Lower limits on $B$ for companions to all A-type (solid line) and G-type (red dashed line) main sequence primaries listed in the WDS, which also have Hipparcos parallax measurements. Companions with $B \geq 1$ cannot be bound, whilst those with $B<1$ may or may not be bound depending on $z$ and $\dot{z}$.}
  \label{fig: wds_B_values}
\end{figure}

For a second example we demonstrate how the simple $B < 1$ criterion may be used to identify unbound companions, by applying the method to visual binaries in the Washington Double Star Catalogue (WDS). We considered all A- and G-type main sequence (MS) primaries listed in the WDS with parallax measurements from Hipparcos \citep{Perryman97}, with observations spanning at least a year; this equates to 1887 visual companions to 1548 A stars and 2352 companions to 2039 G stars. We wished to find a lower limit on the fraction of these with visual companions listed in the catalogue but which simple analysis shows cannot be bound, by calculating a lower limit on $B$ for each one. From Equation \ref{eq: B}, $B$ is minimised if $R$ and $V$ are minimised and $\mu$ is maximised. For each companion we found a lower limit on $V$ from the difference between its projected position relative to the primary at the first and last observational epochs, and took $R$ to be the smaller of the projected separations at these epochs. We used Mamajek's online tables\footnote{\url{http://www.pas.rochester.edu/~emamajek/EEM_dwarf_UBVIJHK_colors_Teff.txt}} \citep{Pecaut13} to infer each primary's mass from its spectral type and, assuming the primary to be more massive than any companions, found an upper limit on $\mu$ by assuming both stars to be the same mass. We combine these to give a lower limit on $B$, and show the distribution of these $B$ values on Figure \ref{fig: wds_B_values}. We find that of all visual companions to A-type MS stars listed in the WDS, at least 60 per cent (1104 objects) cannot be bound. The value is 50 per cent (1264 objects) for G-type MS primaries. The actual fractions of companions that cannot be bound will be higher, as we used conservative assumptions for our analysis. A more thorough analysis could estimate the companion's mass from its V-band magnitude (assuming it is on the main sequence and at the same distance as the primary), or could utilise all available astrometry rather than just the first and last measurements. However we did not do these here because our aim was simply to illustrate the technique using conservative assumptions, to show how a simple evaluation of $B$ can be used without the need for a detailed orbital analysis.

\section{Additional $z$ and $\dot{z}$ constraints: an application to Fomalhaut}
\label{sec: physically_motivated_priors}

\noindent The value of Figure \ref{fig: element_contours} is also evident when there are other physical reasons why $z$ and $\dot{z}$ may be constrained. If regions of $z$, $\dot{z}$ space can be excluded for certain scenarios then the resulting allowed ranges of orbital elements can be easily read off the contour plots, as well as the degeneracies between them. For example, for Fomalhaut there is additional information on the geometry of the system from images of its debris disc (for this paper we consider only the well resolved outer disc, not the warm debris component found by \citealt{Stapelfeldt04}). Quite how Fomalhaut b is related to the disc, and also how the companion acquired such a high eccentricity, are the subjects of ongoing debate (e.g. \citealt{Lawler14, Tamayo14, Faramaz15}). In this section we consider how some simple hypotheses about Fomalhaut b's orbit in relation to this disc affect the allowed ranges of its orbital elements.

Fomalhaut's outer disc is narrow, hence is well approximated by a single ring of debris. In our coordinate system, a ring tracing the centre of the disc has orbital elements $a_{\rm disc} = 141.8$ au, $e_{\rm disc} = 0.1$, $i_{\rm disc} = 66.1^\circ$, $\Omega_{\rm disc} = 199.2^\circ$ and $\omega_{\rm disc} = 30^\circ$ (\citealt{Kalas13}; also note the degeneracy between $\Omega_{\rm disc}$, $\omega_{\rm disc}$ and $\Omega_{\rm disc} + 180^\circ$, $\omega_{\rm disc} + 180^\circ$). Hence the disc's longitude of pericenter $\varpi_{\rm disc}$ equals $229^\circ$. For Fomalhaut b we find $203^\circ \leq \varpi \leq 304^\circ$, hence (if bound) its pericentre must be reasonably well aligned with that of the disc \textit{regardless} of its line of sight coordinates (as found by \citealt{Kalas13} and \citealt{Beust14}). This alignment is also clear from the $\Omega$ versus $\omega$ panels on Figure \ref{fig: element_distributions}, where most points lie close to the $\varpi = \varpi_{\rm disc}$ line.

\begin{figure}
  \centering
      \includegraphics[width=6.06667cm]{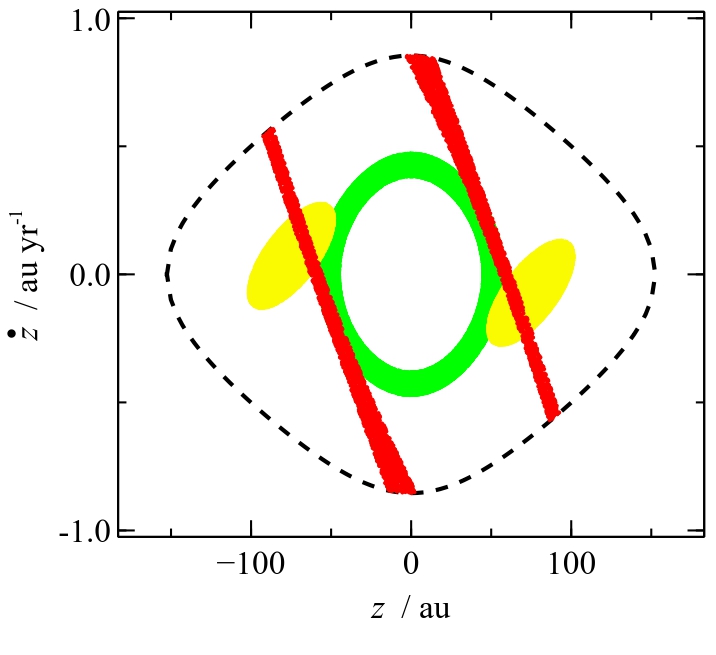}
  \caption{Values of $z$ and $\dot{z}$ allowed by some physically motivated priors for Fomalhaut b. Yellow points: Fomalhaut b orbits within $5^\circ$ of the disc plane. Green: its semi-major axis is within 10 au of the disc's. Red: its orbit passes within 5 au of the middle of the disc when it crosses it in projection. This plot may be overlayed on Figure \ref{fig: element_contours} to understand the element degeneracies on Figure \ref{fig: element_distributions}. The degeneracy between $\Omega_{\rm disc}$, $\omega_{\rm disc}$ and $\Omega_{\rm disc} + 180^\circ$, $\omega_{\rm disc} + 180^\circ$ results in two sets of symmetrical solutions for each prior.}
  \label{fig: z_vz_families}
\end{figure}

\begin{figure}
  \centering
      \includegraphics[width=8cm]{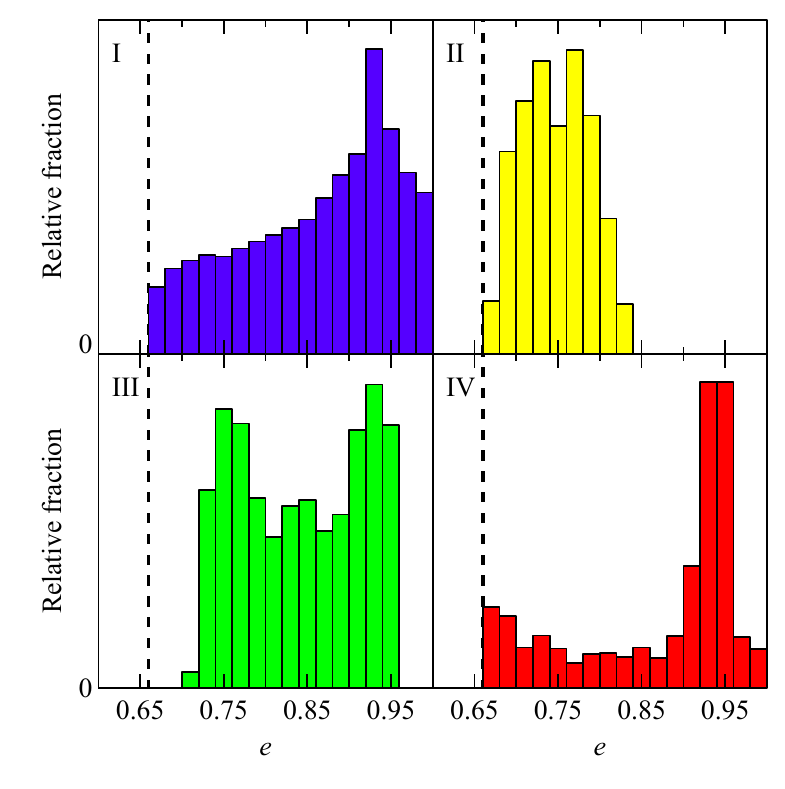}
  \caption{The effect of different physical priors on the distribution of Fomalhaut b's eccentricity solutions. I: uniform $z$ and $\dot{z}$ priors. II: Fomalhaut b orbits within $5^\circ$ of the debris disc plane. III: Fomalhaut b's semi-major axis is within 10 au of the disc's. IV: Fomalhaut b's orbit passes within 5 au of the disc when crossing it in projection in $\sim 50$ years' time. The dashed lines show the lower bound on the eccentricity calculated using the method in Section \ref{sec: distribution_bounds}.}
  \label{fig: e_diff_priors}
\end{figure}

The most obvious hypothesis is that Fomalhaut b is moving in (or close to) the disc plane, as might be expected for a coplanar planetary system (or one in which Fomalhaut b was scattered outwards by a planet coplanar with the disc). If Fomalhaut b orbits in the disc plane then its present day line of sight coordinates are $z = \pm 73$ au and $\dot{z} = \mp 0.062$ au yr$^{-1}$, where the upper or lower signs are taken simultaneously and two solutions arise from the degeneracy in the disc orientation. Relaxing slightly the hypothesis of exact coplanarity so that Fomalhaut b's orbital plane lies within $5^\circ$ of the disc plane (i.e. the angle between the normals of the two orbital planes is smaller than $5^\circ$) then $z$ and $\dot{z}$ must lie in the yellow regions on Figure \ref{fig: z_vz_families}. The yellow points on Figure \ref{fig: element_distributions} show the corresponding orbital solutions, and plot II on Figure \ref{fig: e_diff_priors} shows the resulting distribution of eccentricity solutions. By overlaying Figure \ref{fig: z_vz_families} onto Figure \ref{fig: element_contours}, the allowed ranges of (and degeneracies between) the orbital elements are completely defined. If Fomalhaut b lies within $5^\circ$ of the disc plane it must have 140 au $\leq a \leq$ 320 au, $0.67 \leq e \leq 0.84$ and $214^\circ \leq \varpi \leq 244^\circ$.

We can consider alternative origins for Fomalhaut b, and how these affect its possible orbit. If it originated in the imaged debris disc and evolved onto an eccentric orbit via secular or resonant perturbations from another body, its semi-major axis would be similar to that of the disc (prior III on Figure \ref{fig: e_diff_priors}). Orbits with semi-major axes within 10 au of that of the disc are shown in green on Figures \ref{fig: element_distributions} and \ref{fig: z_vz_families}. We can understand these constraints on $z$ and $\dot{z}$ using the top left plot of Figure \ref{fig: element_contours}, as these points follow an $a \approx 140$ au contour. Again by comparing these $z$ and $\dot{z}$ values to the other plots on Figure \ref{fig: element_contours}, we see that if Fomalhaut b's semi-major axis is within 10 au of that of the disc then it has $0.7 \leq e \leq 0.95$ and $17^\circ \leq i \leq 65^\circ$. The small overlap with the yellow points shows that it is possible for Fomalhaut b to lie in the disc plane \textit{and} have the same semi-major axis as the disc. Larger mutual inclinations are also possible; $\Omega$ and $i$ define the orbital plane, and the fact that many green points lie far from the yellow (coplanar) points on Figure \ref{fig: element_distributions}'s $\Omega$ versus $i$ panel shows that some orbits have high mutual inclinations. The orbital solutions have $203^\circ \leq \varpi \leq 234^\circ$, i.e. an even closer alignment with the disc's pericentre. Pericentre alignment could be expected in this scenario because an unseen object perturbing Fomalhaut b would also perturb the disc, causing alignment of these orbits (although this is not always the case: see \citealt{Beust14}).

Finally we consider a third physically-motivated prior (prior IV on Figure \ref{fig: e_diff_priors}). In $\sim 50$ years' time Fomalhaut b will cross in projection the ring we use to approximate the disc, and we investigate which orbits actually pass through the disc when this occurs. This scenario could be expected if Fomalhaut b originated in the disc, and was later scattered out by another body. Interestingly, this scenario may have a directly observable consequence, since a physical crossing may increase the debris collision rate causing a temporary brightening of the region in scattered light \citep{Kalas13}. The red points on Figures \ref{fig: element_distributions} and \ref{fig: z_vz_families} show orbits passing within 5 au of the three-dimensional disc location that Fomalhaut b will pass in projection. The linear distribution of red points arises because Fomalhaut b's motion will remain roughly linear between its current location and the time when it crosses the disc, so every $z$ value has a corresponding $\dot{z}$ pointing towards the disc (although there are two solutions on Figure \ref{fig: z_vz_families} due to degeneracy in the disc's line of sight position). For Fomalhaut b to cross the disc in $\sim 50 $ years it must be inclined by at least $20^\circ$ to the sky plane, and the allowed ranges of the other elements are unconstrained (beyond the general bounds from Section \ref{sec: distribution_bounds}). In particular note that the orbit need not be coplanar with the disc to pass through this point; high mutual inclinations are also allowed.

The above discussion shows that specific origin scenarios for Fomalhaut b require orbits within restricted ranges of the possible parameter space. It also emphasised the effect of different priors on the distributions of orbital elements. A similar method can be applied to any companion if additional constraints can be placed on its line of sight coordinates.


\section{Limitations of the method}
\label{sec: limitations}

\noindent The method of characterising an imaged companion's orbit using contour plots and bounds on allowed elements has several benefits; it is quick, prior independent, probes the full region of parameter space, allows readers to draw possible orbits themselves, makes degeneracies between elements readily apparent and is particularly useful if physical arguments can further constrain $z$ and $\dot{z}$ (see Section \ref{sec: physically_motivated_priors}). However the method also has limitations, so should be used in conjunction with (not instead of) other techniques.

Firstly the method considers only a single position and velocity, so it does not account for orbital curvature. For a companion with observed curvature, the method is still valid at each instantaneous position and velocity along its path. However many of the resulting orbital solutions would be invalid, because the curvature would rule them out. That only one position and velocity is considered is also a limitation even if curvature is not observed, as linear motion spanning a long enough arc is more constraining than an instantaneous position and velocity. An example is PZ Tel B, which has roughly linear motion but whose separation has doubled over a few years. This extra constraint rules out some of the solutions of our simple analysis, hence the possible orbital parameters found by \citet{Ginski14} cover a smaller range than our solutions. Finally, incorporating additional constraints into our method can sometimes be difficult. For example, our method could identify the minimum possible eccentricity from astrometry data, but additional radial velocity measurements could rule out the combination of line of sight coordinates required to give this minimum eccentricity. This may not be too large a problem though; our method is most applicable to companions far from their host stars, thus radial velocity data may not impose significant constraints.

All of the above limitations are accounted for by MCMC, but that method has its own disadvantages and does not have all the benefits of our method. Therefore the two techniques should both be used in tandem. In particular our method could be used as soon as orbital motion is detected, to quickly calculate the allowed bounds on orbital elements. Once curvature is observed, an MCMC approach may be more appropriate.


\section{Conclusions}
\label{sec: conclusions}

\noindent Imaging a companion over a short orbital arc yields only four of the six coordinates required for a unique orbital solution. MCMC analysis is often used to generate probability distributions associated with the orbital elements, but despite the advantages of such methods, the distributions are influenced by the choice of priors, may not probe the full parameter space and may be misleading if degeneracies are not accounted for. We suggest that an alternative way to characterise possible orbits is to show how the elements differ as functions of the assumed line of sight coordinates, thus allowing a reader to extract potential orbits themselves. Additionally we provide a simple criterion a companion must satisfy to be bound, as well as general constraints on its semi-major axis, eccentricity, inclination, true anomaly and longitude of pericentre. Quoting such bounds would be useful when characterising possible orbits in a prior-independent manner. Our methods are complimentary to MCMC analysis, both techniques having their advantages and disadvantages. In particular our methods may be quickly applied on detection of a companion's orbital motion, without the need for numerical analyses.

As an example we apply our methods to visual companions of A- and G-type main sequence stars in the WDS, and find that about 50 per cent cannot be bound. We also calculate bounds on the elements of several imaged sub-stellar companions, and our results agree with literature values found using more complicated analyses. Finally we considered the effect of physically motivated priors on the allowed orbit of Fomalhaut b; stipulating that its orbit must lie within $5^\circ$ of the debris disc plane means it must have 140 au $\leq a \leq$ 320 au and $0.67 \leq e \leq 0.84$. Alternatively if its semi-major axis is within 10 au of that of the disc then $0.7 \leq e \leq 0.95$, and if its orbit intercepts the disc when passing it in projection then $i > 20^\circ$.


\section{Acknowledgements}
\noindent We thank our referee, Herv\'{e} Beust, who helped us to improve the quality of this paper. TDP acknowledges the support of an STFC studentship, and MCW and GMK are grateful for support from the European Union through ERC grant number 279973. This work made use of the Washington Double Star Catalogue and the VizieR catalogue access tool.



\appendix

\section{Deriving $R$, $V$ and $\varphi$ from orbital data}
\label{app: RV_from_obs}

\noindent The following equations give $R$, $V$ and $\varphi$ in terms of sky separation angle $S$ and position angle $P$ of the binary at two epochs (subscripts 1 and 2), separated by time $\Delta t$. Trivially $R = S_1 d$ and $V = \frac{d}{\Delta t} \sqrt{S_1^2 - 2 S_1 S_2 \cos(P_2 - P_1) + S_2^2}$, and $\varphi$ is given by

\begin{equation}
\cos \varphi = \frac{S_2 \cos(P_2 - P_1) - S_1}{\sqrt{S_1^2 - 2 S_1 S_2 \cos(P_2 - P_1) + S_2^2}}.
\end{equation}

\noindent Note we define $0 \leq \varphi \leq 180^\circ$. Alternatively, in terms of the companion's Northern $N$ and Eastern $E$ sky offsets from the star, we have $R = d\sqrt{N_1^2 + E_1^2}$, $V = \frac{d}{\Delta t} \sqrt{(N_2 - N_1)^2 + (E_2 - E_1)^2}$ and

\begin{equation}
\cos \varphi = \frac{N_1 N_2 + E_1 E_2 - N_1^2 - E_1^2}{\sqrt{N_1^2+E_1^2}\sqrt{(N_2 - N_1)^2 + (E_2 - E_1)^2}}.
\end{equation}


\section{Orbital elements from Cartesian coordinates}
\label{app: element_equations}

The orbital elements of a binary component at position $\mathbf{r} = (x,y,z)$ and velocity $\mathbf{v} = (\dot{x},\dot{y},\dot{z})$ relative to the other (with $\mathbf{h} \equiv \mathbf{r} \times \mathbf{v}$) are

\begin{equation}
 a = \left(\frac{2}{r} - \frac{v^2}{\mu} \right)^{-1},
\label{eq: a}
\end{equation}

\begin{equation}
 e = \sqrt{1- \frac{h^2}{a \mu}},
\label{eq: e}
\end{equation}

\begin{equation}
 \cos i = \frac{\mathbf{h}.\mathbf{\hat{z}}}{h},
\label{eq: i}
\end{equation}

\begin{equation}
 \sin \Omega = \frac{\pm \mathbf{h}.\mathbf{\hat{x}}}{h \sin i} {\text {\;\;\;\; and \;\;\;\;}} \cos \Omega = \frac{\mp \mathbf{h}.\mathbf{\hat{y}}}{h \sin i},
\label{eq: O}
\end{equation}

\begin{equation}
 \sin\theta = \frac{z}{r\sin i} {\text {\;\;\;\; and \;\;\;\;}} \cos\theta = \sec\Omega\left(\frac{x}{r} + \frac{z}{r}\frac{\sin\Omega}{\tan i} \right),
\label{eq: theta}
\end{equation}

\begin{equation}
 \sin f =\frac{a(1-e^2)}{he}\dot{r} {\text {\;\;\; and \;\;\;}} \cos f = \frac{a(1-e^2)-r}{re},
\label{eq: f}
\end{equation}

\noindent where $\theta \equiv \omega + f$,

\begin{equation}
\dot{r} = {\rm sgn}(\mathbf{r}.\mathbf{v}) \sqrt{v^2 - \frac{h^2}{r^2}}
\label{eq: rdot}
\end{equation}

\noindent and the upper signs in Equation \ref{eq: O} are taken if $\mathbf{h}.\mathbf{\hat{z}} > 0$ and the lower if $\mathbf{h}.\mathbf{\hat{z}} < 0$ \citep{Murray and Dermott}. Restating the eccentricity equation in terms of the dimensionless parameters $B$ and $\varphi$ yields

\begin{multline}
e = \biggr[1-4B\left(\rho^2-2\rho\nu\cos\varphi + \nu^2 + \sin^2\varphi\right) \\ \times \left((1+\rho^2)^{-1/2}-B(1+\nu^2)\right)\biggr]^{1/2},
\label{eq: e full}
\end{multline}

\noindent where $\rho \equiv z/R$ and $\nu \equiv \dot{z}/V$.

\label{lastpage}

\end{document}